\documentclass[12pt]{article}
\usepackage[T2A]{fontenc}
\usepackage[utf8]{inputenc}
\usepackage[english]{babel}
\usepackage{amssymb}
\usepackage{amsmath}
\usepackage{graphicx}
\usepackage{caption}
\usepackage[small,sc,center]{titlesec}
\usepackage{indentfirst}
\usepackage{siunitx}

\newcommand{\nc}{\newcommand}
\nc{\fov}{f_\mathrm{ov}}
\nc{\itp}{i_\mathrm{TP}}
\nc{\lhe}{L_\mathrm{He}}
\nc{\mzams}{M_\mathrm{ZAMS}}
\nc{\rph}{R_\mathrm{ph}}
\nc{\tev}{t_\mathrm{ev}}
\begin{document}

\captionsetup[figure]{labelfont={bf},name={Fig.},labelsep=period}

\begin{center}
\textbf{A model of the Mira--type star T UMi}

\vskip 3mm
\textbf{\quad Yu. A. Fadeyev\footnote{E--mail: fadeyev@inasan.ru}}

\textit{Institute of Astronomy, Russian Academy of Sciences, Pyatnitskaya ul. 48, Moscow, 119017 Russia} \\

Received February 6, 2018
\end{center}

\textbf{Abstract} ---
Stellar evolution calculations were carried out from the main sequence
to the final stage of the asymptotic giant branch for stars with initial
masses $1M_\odot\le\mzams\le 2M_\odot$ and metallicity $Z=0.01$.
Selected models of evolutionary sequences were used as initial conditions
for solution of the equations of radiation hydrodynamics and time--dependent
convection describing radial stellar pulsations.
The study was aimed to construct the hydrodynamic models of Mira--type
stars that show the secular decrease in the pulsation period $\Pi$ commenced
in 1970--th at $\Pi=315$ day.
We show that such a condition for the period change is satisfied with
evolutionary sequences $1M_\odot\le\mzams\le 1.2M_\odot$ and
the best agreement with observations is obtained for $\mzams=1.2M_\odot$.
The pulsation period reduction is due to both the stellar radius decrease
during the thermal pulse of the helium burning shell
and mode switch from the fundamental mode to the first overtone.
Theoretical estimates of the fundament parameters of the star at the onset
of pulsation period reduction are as follows:
the mass is $M=0.93M_\odot$, the luminosity is $L = 4080L_\odot$,
and the radius is $R=220R_\odot$.
The mode switch occurs 35 years after the onset of period reduction.


\newpage
\section{introduction}

The long--period $o$~Cet type pulsating variable stars (Miras) are at
a late--stage of their evolution and on the Hertzsprung--Russel diagram (HRD)
they populate the asymptotic giant branch (AGB).
The energy source of Miras are the reactions of thermonuclear burning
of hydrogen and helium in the shells surronding the degenerate carbon--oxygen
core.
The energy generation in the helium burning shell is thermally unstable
and cyclically increases up to values several orders of magnitude higher
than the stellar luminosity (Schwarzschild and H\"arm 1965; Weigert 1966).
The mean time interval between helium flashes depends on the stellar
mass and ranges from $10^4$ to $10^5$ yr (Herwig 2000; Weiss, Ferguson 2009),
whereas the time scale of the changes in the helium burning shell luminosity
is $\le 10^3$ yr (Wood, Zarro, 1981; Boothroyd, Sackmann 1988).
One of the consequences of thermal instability is secular period change observed
in some Miras because during the short--term variation of the helium
burning shell luminosity the stellar radius varies by as much as a factor
of three (Wood, Zarro, 1981; Fadeyev 2016, 2017).

In understanding the nature of Miras there are still uncertainties so that
results of evolutionary computations should be corroborated by tests
based on observations of selected variable stars of this type.
One of such tests is based on the stellar pulsation theory and
allows us to determine the fundamental parameters of the star
(the mass $M$ and the radius $R$) using the observational estimates
of the pulsation period $\Pi$ and the rate of period change $\dot\Pi$.
Efficiency of such an approach was demonstarted in analysis of
cepheid pulsations for $\alpha$~UMi (Fadeyev 2015a) and SZ~Tau (Fadeyev 2015b).
Unfortunately, observational estimates of $\Pi$ and $\dot\Pi$ are
insufficient in the case of Miras because on the HRD the region of AGB
is populated by stars with various initial masses
($1M_\odot\le\mzams\le 7M_\odot$)
and the pulsation period of each AGB star changes in the wide range
due to both thermal pulses and monotonious decrease of the stellar mass
because of the strong stellar wind.

The mira--type star T~UMi is a rare exception which allows us to obtain
an another relationship between observational properties of the star
and its evolutionary status.
This red giant of spectral type M5.5e (Keenan 1966) was pulsating
with period $\Pi\approx 315$ day before 1970s but later the period
commenced its rapid shortening (G\'al, Szatm\'ary 1995) and in 2008
was nearly 114 day (Uttenthaler et al. 2011).
Period decrease observed in T~UMi is thought to be due to the thermal pulse
of the helium shell source
(Mattei, G. Foster 1995; Whitelock 1999; Szatm\'ary et al. 2003).

In our previous work (Fadeyev 2017), we investigated hydrodynamic models
of stars with initial masses $2M_\odot\le\mzams\le 5M_\odot$
on the stage of thermal instability of the helium burning shell (TP--AGB).
In particular, it was shown that reduction of the pulsation period in
T~UMi cannot be explained by models with initial mass $\mzams>2M_\odot$.
The aim of this study is to construct a model of the Mira--type star T~UMi
using the detailed grid of evolutionary sequences with intial masses
ranging within $1M_\odot\le\mzams\le 2M_\odot$.
Similar to our previous works the selected models of evolutionary
sequences are used as initial conditions in solution of the Cauchy problem
for equations of radiation hydrodynamics and time--dependent convection
describing radial stellar oscillations.
Methods of the solution are discussed in our earlier papers (Fadeyev 2016; 2017).

\section{Evolutionary sequences of TP--AGB stars}

Stellar evolution calculations were carried out from the zero--age main sequence
(ZAMS) to the final TP--AGB stage when the star leaves the region of red giants
on the HRD.
Initial relative mass abundances of hydrogen and the elements heavier than helium
were assumed to be $X=0.7$ and $Z=0.01$.
The ratio of the mixing length to the pressure scale height is
$\alpha_\mathrm{MLT} =  1.8$.
Evolutionary computations were done with the MESA code version 9575
(Paxton et al., 2011; 2013; 2015).
Details of program usage for AGB stellar evolution computation and the choise of
basic parameters are discussed in our previous papers (Fadeyev 2016; 2017).
Effects of convective overshooting were taken into account according to Herwig (2000)
for the parameter $\fov=0.014$.
The smaller value of $\fov$ in comparison with our earlier calculations
is due to both the smaller initial masses $\mzams$ and the existence of dependence
between the overshooting parameter and the stellar mass (Claret, Torres 2017).

\begin{figure}
\centerline{\includegraphics[width=15cm]{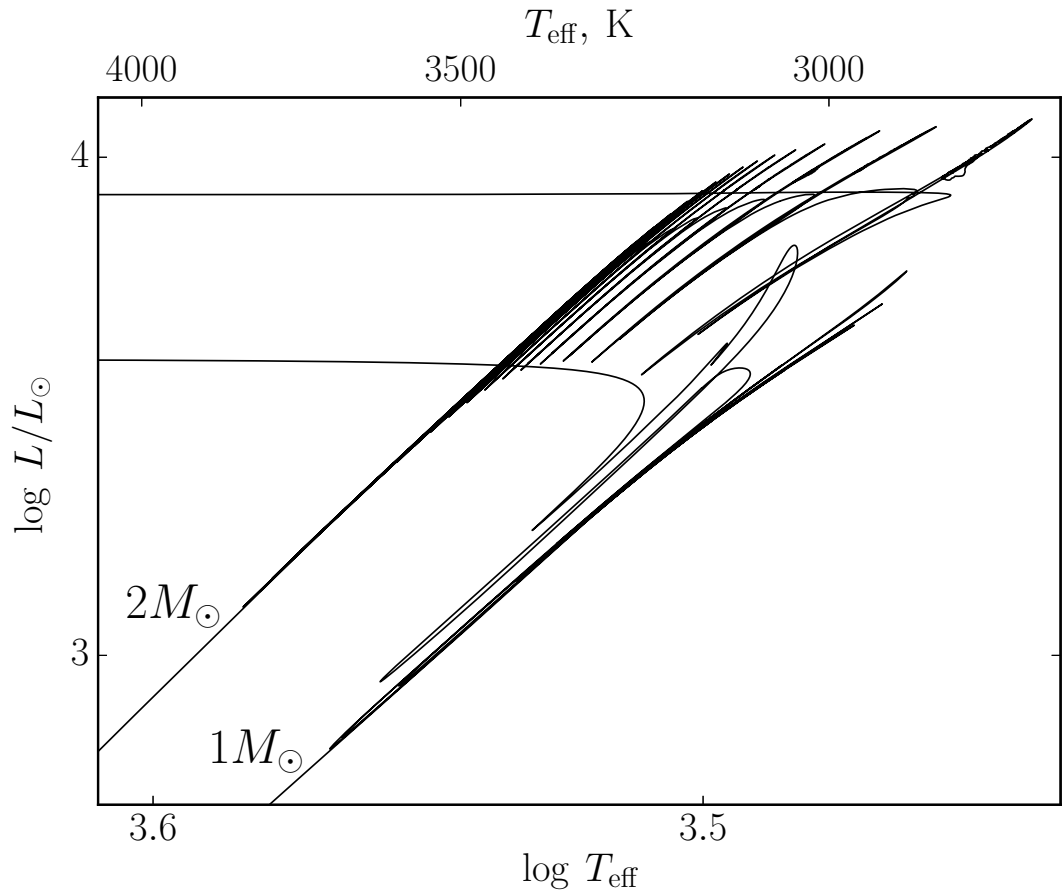}}
\caption{Evolutionary tracks of the stars with initial masses
 $1M_\odot$ and $2M_\odot$ during the AGB stage.}
\label{fig1}
\end{figure}

Fig.~\ref{fig1} shows the evolutionary tracks of stars $\mzams=1M_\odot$ and
$2M_\odot$ during the AGB stage.
Initially evolution on the HRD is accompanied by monotoneous increase of the
luminosity.
This part of the evolutionary track represents the early asymptotic giant branch
(E--AGB) stage when the principal source of the stellar luminosity is the
hydrogen burning shell.
Transition to the stage of thermal instability (TP--AGB) is due to the growth
of the triple $\alpha$--process reaction rates in the helium burning shell.
Helium burning energy release occurs in the form of recurrent pulses.
The number of pulses depends on the stellar mass and increases from 4 for
$\mzams=1M_\odot$ to 19 for $\mzams=2M_\odot$.
Energy release of the helium burning shell ceases and the star leaves the
TP--AGB stage when the total mass of the star reduces to $M\approx 0.63$
for $\mzams=1M_\odot$ and $M\approx 0.83$ for $\mzams=2M_\odot$.
Nearly horizontal parts of the tracks in Fig.~\ref{fig1} correspond to
the post--AGB stage when the stellar effective temperature rapidly increases
at nearly constant luminosity.

It should be noted that both the number of thermal pulses and the
TP--AGB lifetime depend on the mass loss rate relation.
In the present study we assumed that the mass loss rate is given
by relation of Bl\"ocker (1995):
\begin{equation}
\dot M = 4.83\times 10^{-9} \eta_\mathrm{B} (M/M_\odot)^{-2.1}(L/L_\odot)^{2.7} \dot M_\mathrm{R} ,
\end{equation}
where
\begin{equation}
\dot M_\mathrm{R} = 4\times 10^{-13} \eta_\mathrm{R} (L/L_\odot)(R/R_\odot) (M/M_\odot)^{-1}
\end{equation}
-- is the Reimers (1975) formula, and the parameters are
$\eta_\mathrm{R}=0.5$ and $\eta_\mathrm{B}=0.05$.

\begin{figure}
\centerline{\includegraphics[width=15cm]{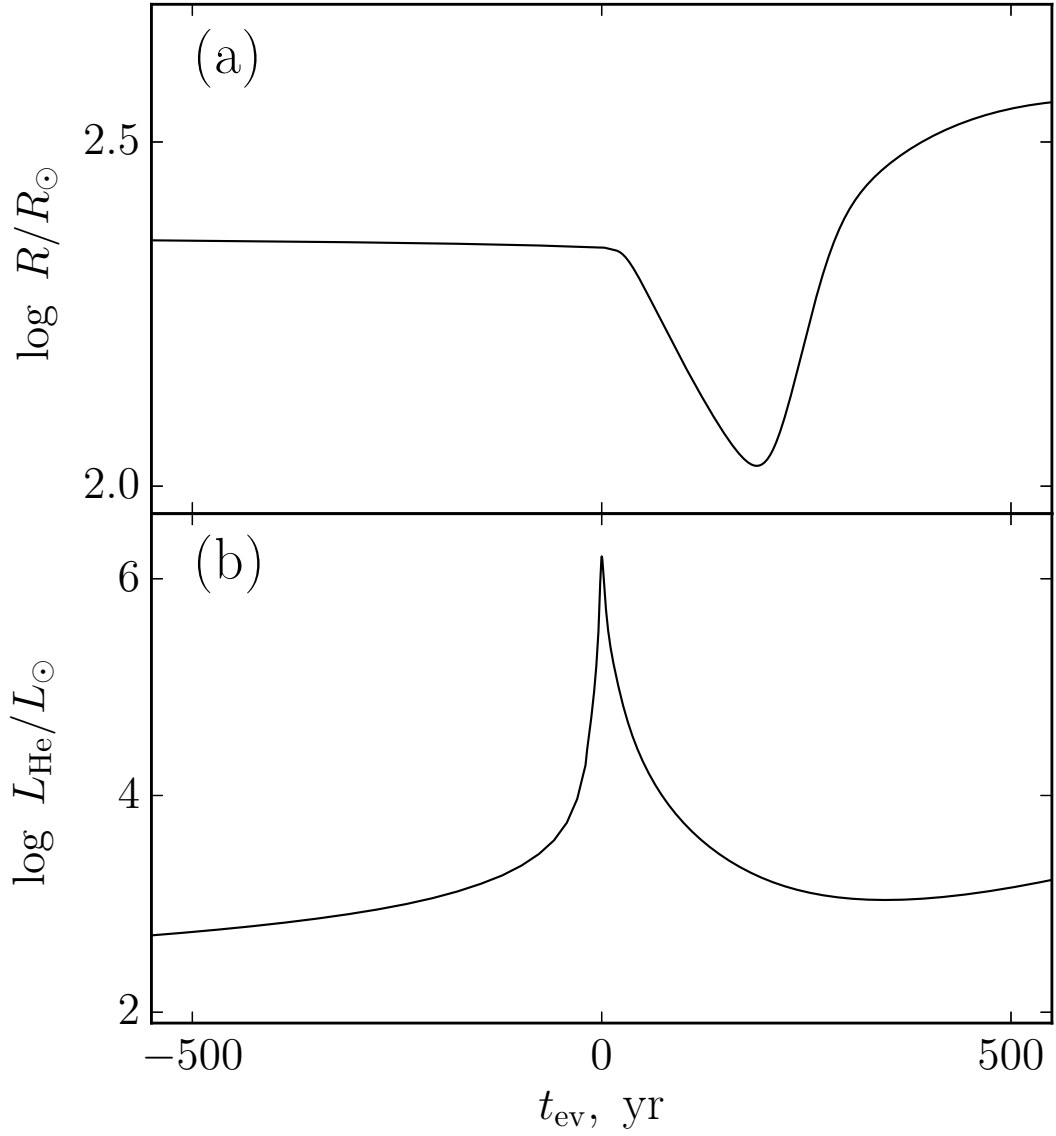}}
\caption{Temporal dependences of the stellar radius $R$ (a) and
the helium burning shell luminosity $\lhe$ (b) during the 4--th thermal pulse
of the evolutionary sequence $\mzams=1.2M_\odot$.}
\label{fig2}
\end{figure}

Fig.~\ref{fig2} shows the temporal dependences of the stellar radius $R$
and the luminosity of the helium burning shell $\lhe$ in the vicinity of the
fourth thermal pulse peak.
For the sake of convenience the evolutionary time $\tev$ is set to zero
at maximum of $\lhe$.
As clearly seen in Fig.~\ref{fig2}, the onset of the stellar radius decrease
coincides with maximum of $\lhe$ within several dozen years.
To a first approximation the radius of the star is related to the period of radial
pulsations by $\Pi\propto R^{3/2}$, therefore without significant loss of accuracy
we can assume that the onset of period decrease coincides with the maximum of $\lhe$.

\begin{figure}
\centerline{\includegraphics[width=15cm]{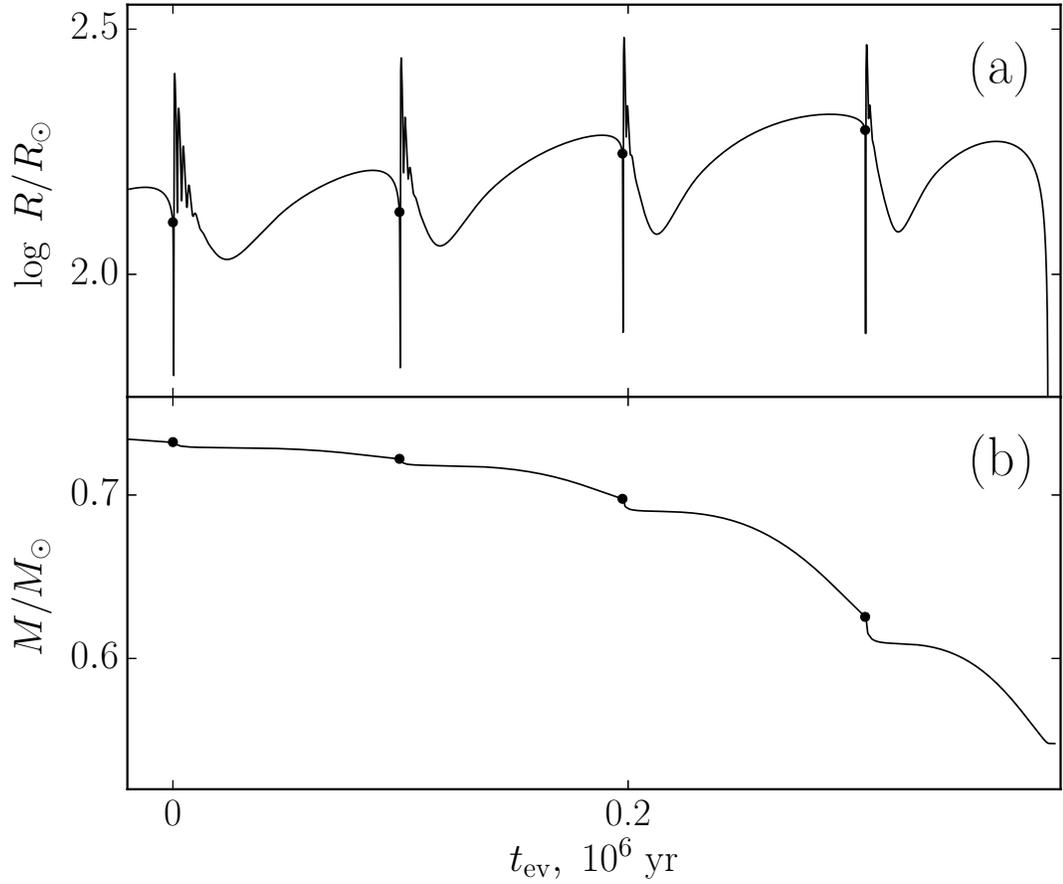}}
\caption{Evolutionary changes of the radius $R$ (a) and the mass $M$ (b)
of the star during the TP--AGB stage for the evolutinary sequence
$\mzams=1M_\odot$.
Filled circles indicate the maxima of $\lhe$.
The evolutionary time $\tev$ is set to zero at the first maximum of $\lhe$.}
\label{fig3}
\end{figure}

\begin{figure}
\centerline{\includegraphics[width=15cm]{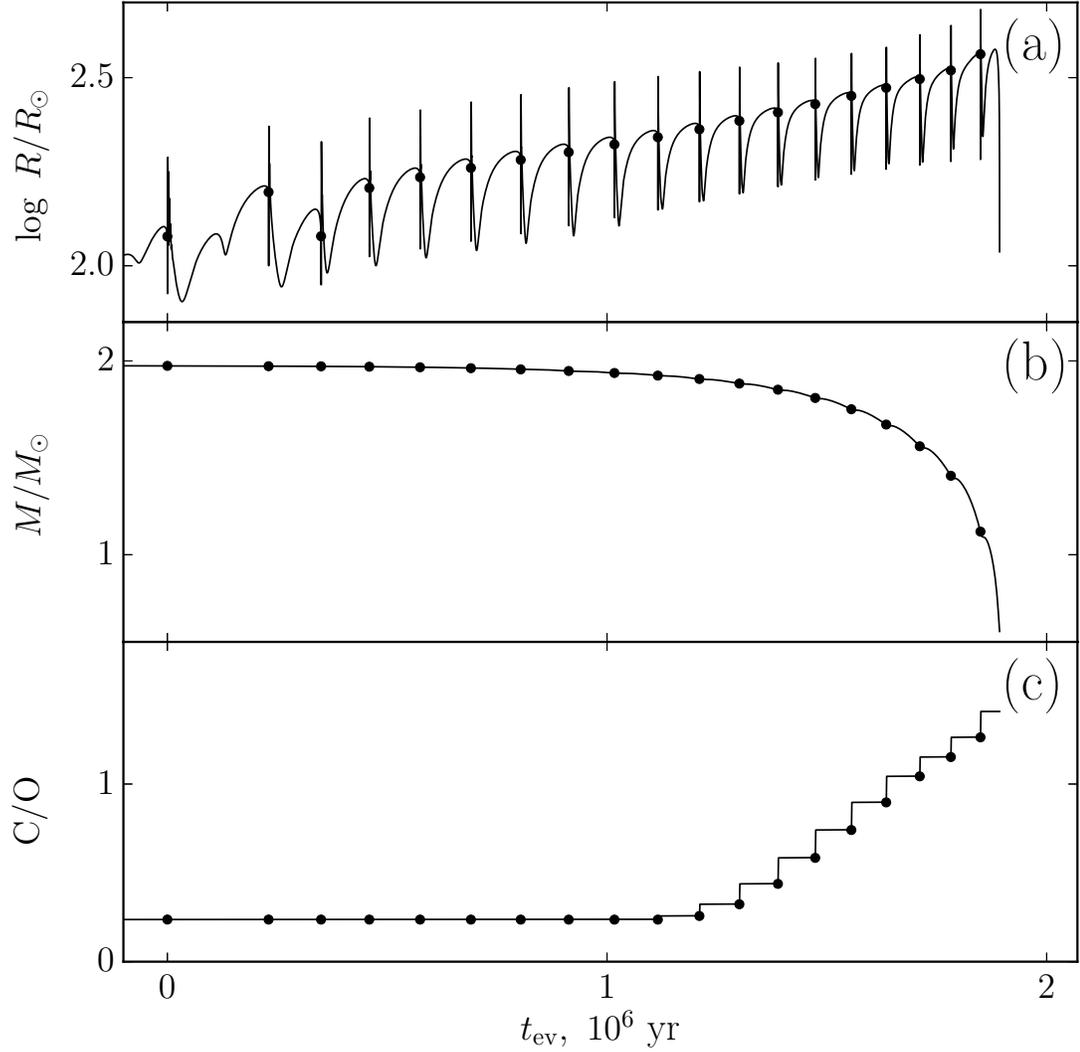}}
\caption{Evolutionary changes of the radius $R$ (a), the mass $M$ (b)
and the surface carbon to oxygen mass fraction ratio $\mathrm{C}/\mathrm{O}$
during the TP--AGB stage for the evolutionary sequence $\mzams=2M_\odot$.
Filled circles indicate the maxima of $\lhe$.
The evolutionary time $\tev$ is set to zero at the first maximum of $\lhe$.}
\label{fig4}
\end{figure}

Evolutionary changes of the stellar radius $R$ and mass $M$ during the TP--AGB
stage are shown in Fig.~\ref{fig3} for the evolutionary sequence $\mzams=1M_\odot$.
The ratio of the carbon to oxygen mass fractions at the stellar surface almost
does not change during the whole TP--AGB stage and is $\mathrm{C}/\mathrm{O}\approx 0.29$.
Fig.~\ref{fig4} shows the plots of the stellar radius and the stellar mass
as a function of evolutinary time $\tev$  for the evolutionary sequence
$\mzams=2M_\odot$.
The bottom panel of Fig.~\ref{fig4} shows also the plot of $\mathrm{C}/\mathrm{O}$.
Filled circles in Figs.~\ref{fig3} and \ref{fig4} indicate the evolutionary
models with peak values of $\lhe$ that were selected as initial conditions
in solution of the equations of hydrodynamics describing stellar pulsations.
It should be noted that selection of evolutionary models for $\mzams=2M_\odot$
is resticted by condition $\mathrm{C}/\mathrm{O} < 1$
because T~UMi belongs to oxygen--rich Miras.

\section{nonlinear stellar pulsations}

As can be clearly seen in Figs.~\ref{fig3} and \ref{fig4}, the stellar radius
$R_\star$ at the maximum of $\lhe$ increases each thermal pulse.
The only exception is the third thermal pulse of the evolutionary sequence
$\mzams=2M_\odot$ when the condition of monotonic increase of $R_\star$
is not fulfilled.
Results of our computations show that the only exception in monotonic
increase of $R_\star$ as a function of the thermal pulse index $\itp$
occurs almost in each evolutionary sequence.
In particular, the condition of monotonic increase is not fulfilled for $\itp=2$
($1.1M_\odot\le\mzams\le 1.6M_\odot$) or for $\itp=3$
($1.8M_\odot\le\mzams\le 2M_\odot$).
Bearing in mind these rare exceptions we try to find the hydrodynamic model
with pulsation period $\Pi=315$ day assuming that $R_\star$ monotonically
increases with $\itp$.

\begin{figure}
\centerline{\includegraphics[width=15cm]{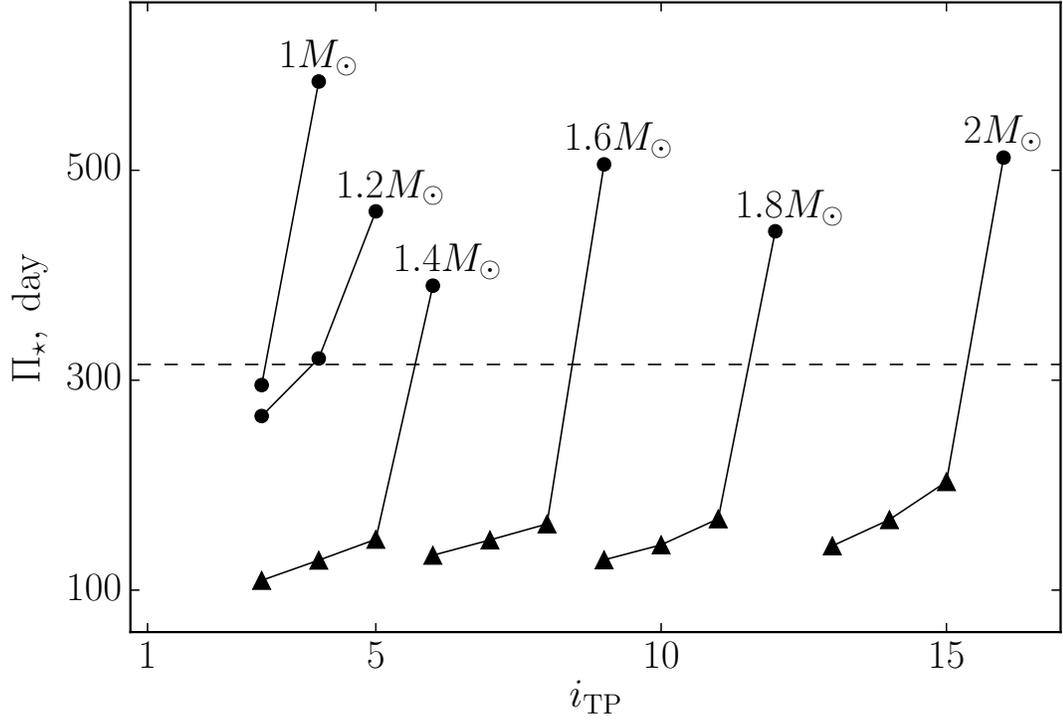}}
\caption{The pulsation period $\Pi_\star$ at maximum of $\lhe$ against
the thermal pulse index $\itp$ for evolutionary sequences from
$1M_\odot$ to $2M_\odot$.
Fundamental mode and first overtone pulsators are shown by filled circles and
filled triangles, respectively.
The horizontal dashed line corresponds to the period $\Pi=315$ day.}
\label{fig5}
\end{figure}

The evolutionary time $\tev$ of the TP--AGB star is sensitive to the initial
mass $\mzams$ therefore to compare the evolutionary sequences it is preferable
to use the index of the thermal pulse $\itp$ as an independent variable.
The period--thermal pulse diagram for evolutionary sequences
$1M_\odot\le\mzams\le 2M_\odot$ is shown in Fig.~\ref{fig5}.
The pulsation periods $\Pi_\star$ at maxima of $\lhe$ are shown by filles
circles and filled triangles for the fundamental mode and first overtone
pulsators, respectively.

Plots in Fig.~\ref{fig5} allow us to conclude that the most appropriate
evolutionary sequences for the model of T~UMi have initial masses
$1M_\odot\le\mzams\le 1.2M_\odot$.
Indeed, only for these sequences one can find two adjacent models
$i$ and $i+1$ pulsating in the same mode with periods
satisfying the condition $\Pi_i < 315~\mbox{day} < \Pi_{i+1}$.
In particular, this condition is fulfilled only for the fundamental mode
pulsations during the third and the fourth termal pulses.
The age of the star is $8.1\times 10^9$, $5.8\times 10^9$ and $4.3\times 10^9$
yr for models $\mzams=1M_\odot$, $1.1M_\odot$ and $1.2M_\odot$, respectively.

\begin{figure}
\centerline{\includegraphics[width=15cm]{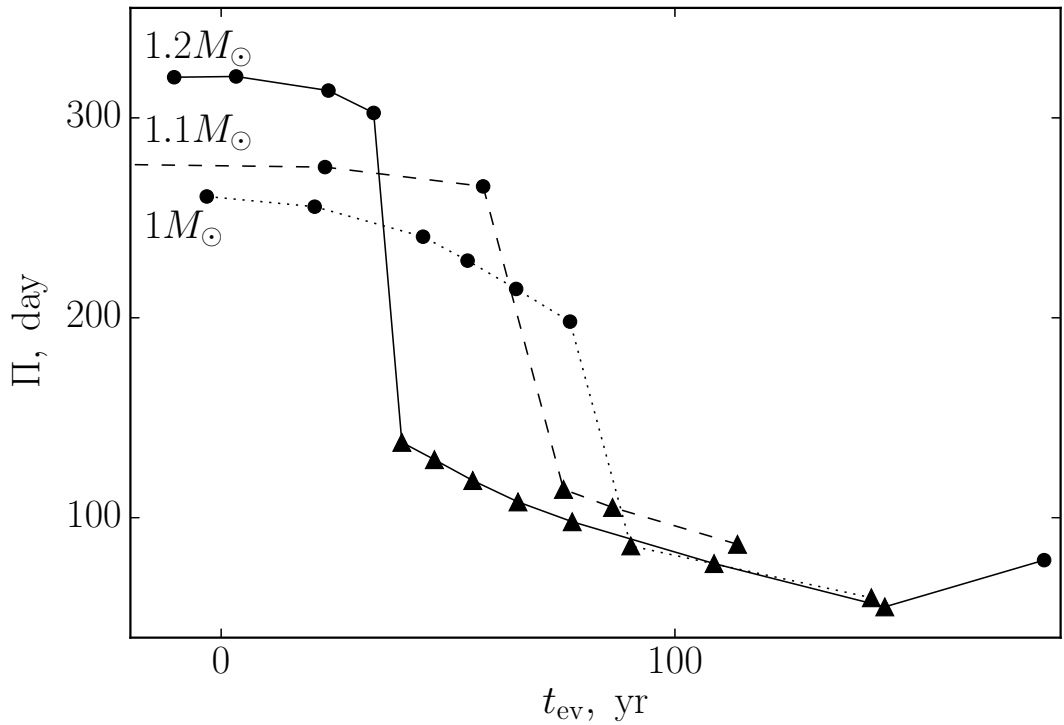}}
\caption{Pulsation period $\Pi$ versus evolutionary time $\tev$ for models
of evolutionary sequences $\mzams=1M_\odot$, $1.1M_\odot$ and $1.2M_\odot$.}
\label{fig6}
\end{figure}

Fig.~\ref{fig6} shows the plots of the pulsation period as a function
of $\tev$ for evolutionary sequences
$\mzams=1M_\odot$, $1.1M_\odot$ and $1.2M_\odot$.
For each sequence the evolutionary time is set to zero at maximum of $\lhe$
and initial pulsation periods are 261, 278 and 320 day.
Of most interest are the hydrodynamic models of the evolutionary sequence
$\mzams=1.2M_\odot$ because the initial value of the pulsation
period $\Pi_\star=321$ day
differ from the observed period of T~UMi by less than 2\%.

In favour of the evolutionary sequence $\mzams=1.2M_\odot$
as the most appropriate model for T~UMi is also the fact that
the time interval $\approx 35$ yr between the onset of period decrease
and the mode switch from the fundamental mode to the first overtone
agrees with observations (Uttenthaler et al. 2011).
As is seen in Fig.~\ref{fig6}, the mode switch occurs later with
decreasing stellar mass and for $\mzams=1M_\odot$
the disagreement between the model and observations
becomes significant ($\approx 80$ yr).

At the onset of period reduction the stellar mass is $M=0.929M_\odot$,
and decrease of the radius and luminosity proceeds for $\approx 180$~yr.
General parameters of the models for this time interval are listed in
the table.
The first column gives the evolutionary time $\tev$ which is set to zero
at the maximum of $\lhe$.
The second and the third columns give the stellar luminosity $L$ and
the radius $R$ of the evolutionary model used as initial conditions in
solution of the equations of hydrodynamics for nonlinear stellar oscillations.
The fourth column gives the mean value of the radius of the photosphere
$\langle\rph\rangle$ obtained from the limit cycle hydrodynamic computations.
The ratio of the mean radius of the photosphere to the initial stellar radius
ranges within $1.03\le\langle\rph\rangle/R\le 1.1$ due to nonlinearity
of stellar pulsations.
In the last three columns we give the pulsation period $\Pi$,
the pulsation constant $Q$ and the order of the pulsation mode $k$.
It should be noted that $Q$ was calculated from the period--mean density
relation with the mean radius of the photosphere $\langle\rph\rangle$.

\begin{figure}
\centerline{\includegraphics[width=15cm]{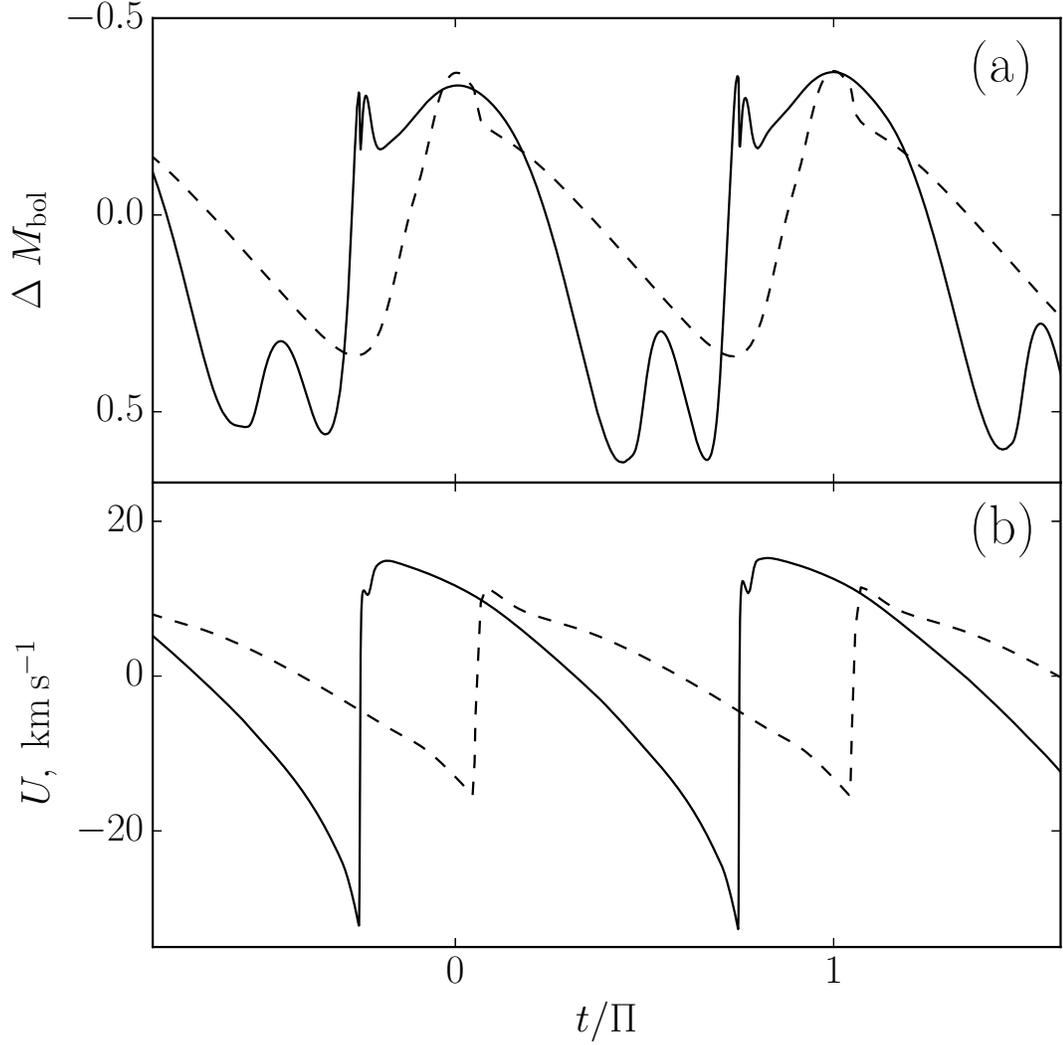}}
\caption{The bolometric light curve $\Delta M_\mathrm{bol}$ (a)
and the surface gas flow velocity $U$ (b) of the Mira--type star T~UMi model
at the onset of the period reduction $\tev=0$ (solid lines) and after
mode switch from the fundamental mode to the first overtone in $\tev=40$ yr (dashed lines).}
\label{fig7}
\end{figure}

Fig.~\ref{fig7} shows the bolometric light and the surface velocity curves
for two hydrodynamic models of the evolutionary sequence $\mzams=1.2M_\odot$.
The first of these models corresponds to the onset of period reduction
($\tev=0$), whereas the second one corresponds to $\tev=40$ yr when
the star became the first overtone pulsator.
The second model is by $0.08$ mag fainter in comparison with the first model.
Smaller amplitude of the light curve is due to mode switch
from the fundamental mode to the first overtone.

\section{conclusions}

Presented above results of stellar evolution and stellar pulsation
calculations for red giants with initial measses from 1 to $2M_\odot$
extend the grid of evolutinary and pulsational models for more massive
($2M_\odot\le\mzams\le 5M_\odot$) AGB stars (Fadeyev 2016; 2017).
Agreement of the family of hydrodynamic models $\mzams=1.2M_\odot$
with observations of the Mira--type star T~UMi corroborates
the stellar evolution computations that provided us with initial conditions
required for solution of the Cauchy problem.
A cause of insignificant disagreement between the theory and observations seems
to be due to the fact that T~UMi is at the beginning of its TP--AGB stage
because the outer layers of the star are still oxygen--rich
without dredgedup material traced by radio--active technetium
(Uttenthaler et al. 2011).
Therefore, uncertainties accompanying evolutionary computations of TP--AGB
stars (extended convective mixing due to overshoot and mass loss rates)
has not played yet significant role in evolutionary and hydrodynamic models.

The best agreement of our evolutionary and hydrodynamic computations
with observations was obtained for the star age $\approx 4.3\times 10^9$ yr.
Therefore, the initial metallicity $Z=0.01$ of the model of T~UMi
seems to be somewhat underestimated.
It should be noted that variation of metallicity within
$0.01\le Z\le 0.02$ do not affect perceptibly both the pulsational instability
and the pulsation period because oscillarions are excited in the
hydrogen ionization zone.

Observations of the Mira--type star T~UMi during near decades are
of great importance because extention of the time interval
for comparison of the theory with observations will allow us to
obtain further constraints on the basic parameters and conclusions
of the stellar evolution theory.

This research was supported by the Basic Research Program P--28
of the Presidium of the Russian Academy of Sciences.

\newpage
\section*{references}

\begin{enumerate}
\item T. Bloecker, Astron. Astrophys. \textbf{297}, 727 (1995).

\item A.I. Boothroyd and I.-J. Sackmann, Astrophys. J. \textbf{328}, 632 (1988).

\item A. Claret and G. Torres), Astrophys. J. \textbf{849}, 18 (2017).

\item Yu.A. Fadeyev, MNRAS \textbf{449}, 1011 (2015а).

\item Yu.A. Fadeyev, Pis'ma Astron. Zh., \textbf{41}, 694 (2015b)
       [Astron.Lett. \textbf{41}, 640 (2015b)].

\item Yu.A. Fadeyev, Pis'ma Astron. Zh. \textbf{42}, 731 (2016)
      [Astron. Lett. \textbf{42}, 665 (2016)].

\item Yu.A. Fadeyev, Pis'ma Astron. Zh., \textbf{43}, 663 (2017)
       [Astron.Lett. \textbf{43}, 602 (2017)].

\item J. G\'al and K. Szatm\'ary, Astron. Astrophys. \textbf{297}, 461 (1995).

\item F. Herwig, Astron. Astrophys. \textbf{360}, 952 (2000).

\item P.C. Keenan, Astropys. J. Suppl. Ser. \textbf{13}, 333 (1966).

\item J.A. Mattei and G. Foster, J. Am. Assoc. Var. Star Observ. \textbf{23}, 106 (1995).

\item B. Paxton, L. Bildsten, A. Dotter, F. Herwig, P. Lesaffre, and F. Timmes,
      Astropys. J. Suppl. Ser. \textbf{192}, 3 (2011).

\item B. Paxton, M. Cantiello,  P. Arras, L. Bildsten,
      E.F. Brown, A. Dotter, C. Mankovich, M.H. Montgomery, et al.,
      Astropys. J. Suppl. Ser. \textbf{208}, 4 (2013).

\item B. Paxton, P. Marchant, J. Schwab, E.B. Bauer,
      L. Bildsten, M. Cantiello, L. Dessart, R. Farmer, et al.,
      Astropys. J. Suppl. Ser. \textbf{220}, 15 (2015).

\item D. Reimers, \textit{Problems in Stellar Atmospheres and Envelopes},
      Ed. by B. Baschek, W.H. Kegel, and G. Traving
      (Springer, New York, 1975), p. 229.

\item M. Schwarzschild and R. H\"arm, Astrophys. J. \textbf{142}, 855 (1965).

\item K. Szatm\'ary, L.L. Kiss, and Zs. Bebesi, Astron. Astrophys. \textbf{398}, 277 (2003).

\item P.A. Whitelock, New Astronomy Reviews \textbf{43}, 437 (1999).

\item A. Weigert, Zeitschrift f\"ur Astrophys. \textbf{64}, 395 (1966).

\item A. Weiss and J.W. Ferguson, Astron. Astrophys. \textbf{508}, 1343 (2009).

\item P.R. Wood and D.M. Zarro, Astrophys. J. \textbf{247}, 247 (1981). 

\item S. Uttenthaler, K. van Stiphout, K. Voet, H. van Winckel, S. van Eck,
      A. Jorissen, F. Kerschbaum, G. Raskin, S. Prins, W. Pessemier, C. Waelkens,
      Y. Fr\'emat, H. Hensberge, L. Dumortier, and H. Lehmann, H.,
      Astron. Astrophys. \textbf{531}, A88 (2011).

\end{enumerate}

\newpage
\begin{table}
\caption{Basic parameters of the model of T~UMi on the stage of stellar radius decrease}
\begin{center}
 \begin{tabular}{rrccrrr}
  \hline
  $\tev$, yr  & $L/L_\odot$ & $R/R_\odot$ & $\langle\rph\rangle/R_\odot$ & $\Pi$, day & $Q$, day & $k$ \\
  \hline
      0.0 &  4077 & 222 & 242 &  320  & 0.0815  & 0 \\
     30.3 &  3905 & 215 & 234 &  303  & 0.0815  & 0 \\
     36.5 &  3791 & 210 & 233 &  138  & 0.0372  & 1 \\
     74.1 &  2934 & 174 & 186 &   98  & 0.0373  & 1 \\
    143.0 &  1816 & 123 & 126 &   55  & 0.0373  & 1 \\
    178.1 &  1514 & 109 & 113 &   79  & 0.0638  & 0 \\
  \hline          
 \end{tabular}
\end{center}
\end{table}
\clearpage

\newpage
\clearpage

\newpage
\clearpage

\newpage
\clearpage

\newpage
\clearpage

\newpage
\clearpage

\newpage
\clearpage

\newpage
\clearpage

\end{document}